\documentclass[prd,showkeys,floatfix,nofootinbib,
               preprint,12pt,
               tightenlines,fleqn]{revtex4} 

\usepackage{amsmath,amssymb,revsymb,graphicx,dcolumn}
\usepackage{leftidx}
\usepackage{slashed}

\usepackage{slashed}

\newcommand{\beq}{\begin{equation}}
\newcommand{\eeq}{\end{equation}}
\newcommand{\beqa}{\begin{eqnarray}}
\newcommand{\eeqa}{\end{eqnarray}}
\newcommand{\bsubeqs}{\begin{subequations}}
\newcommand{\esubeqs}{\end{subequations}}

\usepackage{mwe}
\newcommand{\imineq}[2]{\vcenter{\hbox{\includegraphics[height=#2ex]{#1}}}}

\begin{document}
\hfill KA--TP--05--2017\vspace*{8mm}\newline
\title[]
      {Magnetostatic-field screening induced by small black holes}
\author{Slava Emelyanov}
\email{viacheslav.emelyanov@kit.edu}
\affiliation{Institute for Theoretical Physics,\\
Karlsruhe Institute of Technology (KIT),\\
76131 Karlsruhe, Germany\\}

\begin{abstract}
\vspace*{2.5mm}\noindent
We find within the framework of quantum electrodynamics that there exists screening
effect of static magnetic field that is induced by small evaporating black holes.
\end{abstract}



\maketitle

\section{Introduction}

By this paper, we continue our study of various physical imprints of small black holes in local electromagnetic
phenomena~\cite{Emelyanov-16a,Emelyanov-16b}. The small black holes we have been considering
possess the mass $M$ from the range $10^{10}\,\text{g} \lesssim M \ll 10^{16}\,\text{g}$ which might have
formed through the gravitational collapse at early stages of the universe evolution~\cite{Hawking-1}. This
corresponds to the Hawking temperature $T_H$~\cite{Hawking-2} that is much larger than the electron
rest energy $m_e$. As a consequence, the thermal-like term in the electron 2-point function is not exponentially
suppressed by the Boltzmann factor $\exp(-m_e/T_H)$ as it holds $m_e/T_H \ll 1$. This means that the
electron appears to be effectively massless. This leads to more or less sizeable quantum
effects whenever a small black hole is sufficiently close to a detector.

We employ our recent results obtained in \cite{Emelyanov-17a} to derive the Feynman propagator
$S(x,x')$ of a massless Dirac field in the far-horizon region of a small black hole. This is essentially
given by the ordinary Minkowski propagator $S_M(x,x')$ plus a thermal-like singularity-free correction
$\Delta{S}(x,x')$ decreasing in the spatial infinity as $(r_H/R)^2$, where $R$ is a radial distance to the
black-hole centre and $r_H = 2M$ the size of the event horizon. Although it asymptotically vanishes, the
correction $\Delta{S}(x,x')$ is, nevertheless, physically relevant as being responsible for the evaporation
effect of black holes~\cite{Emelyanov-17a}.

We found in \cite{Emelyanov-16b} that local black-hole manifestations in the electromagnetic
phenomena are characterized by an effective (gauge invariant) photon mass and Debye-like screening
of the electrostatic field of a point-like charge. Therefore, it turns out that the quantum vacuum in the
presence of small black holes shows locally up properties which are usually attributable to a many-particle
system. Specifically, it resembles a hot electron-positron plasma. The purpose of this paper is to show
that there also exists the shielding effect of the magnetostatic field. A similar effect can occur in the hot
electron-positron plasma, but with anisotropic distribution of the constituent particles in momentum
space (like in QCD for the anisotropic quark-gluon plasma~\cite{Kao&Nayak&Greiner,Cooper&Kao&Nayak}).

Throughout this paper the fundamental constants are set to $c=G=k_\text{B} = \hbar = 1$, unless stated
otherwise.

\section{Screening of magnetostatic field}

\subsection{Fermion Feynman propagator}

We derived in \cite{Emelyanov-17a} the scalar 2-point function $W(x,x')$ in the presence of Schwarzschild
black hole formed through the gravitational collapse. This can be exploited to obtain the fermion propagator.
Specifically, the Feynman propagator $S(x,x')$ of a massless fermion in the far-horizon region ($R \gg r_H$)
reads
\beqa\label{eq:fp}
S(x,x') &=& S_M(x,x') + \Delta{S}(x,x')\,,
\eeqa
where
\beqa
S_M(x,x') &\approx& {\int}\frac{d^4p}{(2\pi)^4}\frac{i\slashed{p}}{p^2 + i\varepsilon}\,\exp(-ip\Delta{x})
\eeqa
with $\Delta{x} \equiv x - x'$ and
\beqa\label{eq:fpc}
\Delta{S}(x,x') &\approx& - 2\pi g_R{\int}\frac{d^4p}{(2\pi)^4}\frac{\delta(p^2)}{e^{\beta |p_0|} + 1}
\,\slashed{\bar{p}}\,\exp(-i\bar{p}\Delta{x}) \quad \text{with} \quad
\bar{p}^\mu \;=\; (p_0,p_0\mathbf{n})\,,
\eeqa
where $\beta = 1/T_H$ is the inverse Hawking temperature, $\mathbf{n} \equiv \mathbf{R}/R$ is the radial unit
vector and
\beqa
g_R &\equiv& \frac{27}{16}\Big(\frac{r_H}{R}\Big)^2\,.
\eeqa
It should be emphasised that $\Delta{S}(x,x')$ solves the field equation up to the terms vanishing as $1/R^3$
at spatial infinity. The correction $\Delta{W}(x,x')$ to $W_M(x,x')$ found in~\cite{Emelyanov-17a} satisfies
the scalar field equation in the limit $\mathbf{x}' \rightarrow \mathbf{x}$ only. Therefore, $\Delta{S}(x,x')$ is
a more general result (see Appendix \ref{app:feynman-propagator} for further details).

The fermion stress tensor $\langle \hat{T}_\nu^\mu \rangle$ can be computed by taking its trace with
respect to the spinorial indices and using the equation $\text{tr}\big(\bar{\psi}\gamma_\mu\partial_\nu\psi\big) = 
-\lim_{x' \rightarrow x}\text{tr}\big(\gamma_\mu\partial_\nu S(x,x')\big)$. Making use of $S(x,x')$
given in Eq.~\eqref{eq:fp}, we find the renormalised energy-momentum tensor:
\beqa\label{eq:emt}
\langle \hat{T}_\nu^\mu \rangle &\approx&
\frac{2}{4\pi R^2}{\int\limits_0^{+\infty}}\frac{dp_0}{2\pi}\frac{p_0 \Gamma_{p_0}}{e^{\beta p_0} + 1}
\left[
\begin{array}{cc}
+ 1 & + 1 \\
- 1 & - 1
\end{array}
\right] \quad \text{with} \quad \Gamma_{p_0} \;\equiv\; 27(p_0 M)^2\,,
\eeqa
where the indices $\mu,\nu$ run over $\{t,r\}$ and the rest elements of $\langle \hat{T}_\nu^\mu \rangle$
vanish faster than $1/R^2$ at $R \gg r_H$. This result implies that $S(x,x')$ is a correct expression of the exact
propagator up to terms vanishing faster than $1/R^2$ in the far-horizon region and for points 
satisfying the condition $|r-r'| \ll R$.

\subsection{One-loop vacuum polarisation tensor}

In order to study how the presence of a small black hole can influence the local electro-magnetic phenomena,
one needs to compute the vacuum polarisation tensor $\Pi^{\mu\nu}(k)$. At one-loop approximation, it is given
pictorially by
\beqa\label{eq:vpt}
i\Pi^{\mu\nu}(k) &=& 
\mathbf{\imineq{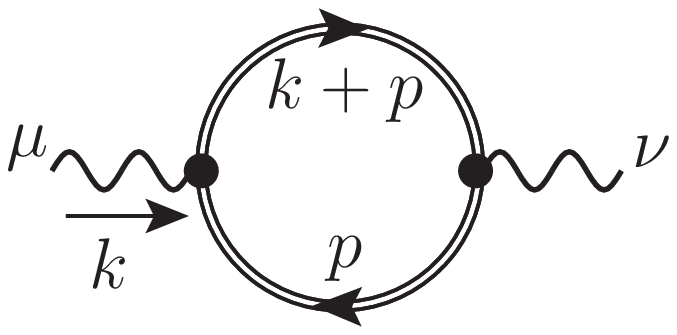}{10}}\hspace{-5mm}\,,
\eeqa
where the double line in the fermion loop refers to the propagator $S(x,x')$ that is composed of the
ordinary part $S_M(x,x')$ and the correction $\Delta{S}(x,x')$ to it. We focus here only on that part
of $\Pi^{\mu\nu}(k)$ which is induced by the presence of a small black hole. This reads
\beqa
\Delta\Pi^{\mu\nu}(k) &=& -4\pi g_Re^2{\int}\frac{d^4p}{(2\pi)^4}\frac{\delta(p^2)}{e^{\beta |p_0|}+1}
\frac{\text{tr}\big(\gamma^\mu \slashed{\bar{p}}\gamma^\nu(\slashed{\bar{p}} + \slashed{k})\big)}
{(\bar{p} + k)^2 + i\varepsilon}\,.
\eeqa
This can in turn be rewritten in terms of the projection tensors $P^{\mu\nu}$ and $Q^{\mu\nu}$
introduced in~\cite{Weldon} as follows:
\beqa
\Delta\Pi^{\mu\nu}(k) &=& \pi_T(k_0,\mathbf{k})P^{\mu\nu} + \pi_L(k_0,\mathbf{k})Q^{\mu\nu}\,,
\eeqa
where we have
\bsubeqs\label{eq:pit-and-pil}
\beqa
\pi_T(k_0,\mathbf{k}) &=& \frac{4g_Re^2}{\pi^2}{\int\limits_0^{+\infty}}\frac{dp\,p^3}{e^{\beta p} + 1}
\frac{4k_0|\mathbf{k}|\cos\theta - (k_0^2 + |\mathbf{k}|^2)(\cos^2\theta + 1)}{(k_0^2 - |\mathbf{k}|^2)^2
- 4p^2(k_0 - |\mathbf{k}|\cos\theta)^2}\,,
\\[1mm]
\pi_L(k_0,\mathbf{k}) &=& \frac{8g_Re^2}{\pi^2}{\int\limits_0^{+\infty}}\frac{dp\,p^3}{e^{\beta p} + 1}
\frac{(k_0^2 - |\mathbf{k}|^2)(\cos^2\theta - 1)}{(k_0^2 - |\mathbf{k}|^2)^2 - 4p^2(k_0 - |\mathbf{k}|\cos\theta)^2}\,
\eeqa
\esubeqs
with $\theta$ being the angle between $\mathbf{k}$ and the radial unit vector $\mathbf{n}$, i.e.
$\cos\theta = \mathbf{k}{\cdot}\mathbf{n}/|\mathbf{k}|$. The integrals
in Eqs.~\eqref{eq:pit-and-pil} are understood as the principal value ones. It should also be stressed out that the
structure of $\pi_T(k_0,\mathbf{k})$ and $\pi_L(k_0,\mathbf{k})$ significantly differs from that in the hot
(isotropic) electron-positron plasma.

In the absence of the black hole, the polarization tensor $\Pi^{\mu\nu}(k)$ has the standard non-trivial
form, $\Pi_M^{\mu\nu}(k)$, and leads to the running effect of the electric charge. This part of the
polarization tensor $\Pi^{\mu\nu}(k)$ starts to reveal itself at the microscopic scale that is of the order of the
Compton wavelength of the electron $\lambda_e \approx 2.4{\times}10^{-12}\,\text{m}$. We are
interested, however, in the low-energy effects which correspond to the length scale of the order of
$1\,\text{m}$ (see below). Thus, we omit $\Pi_M^{\mu\nu}(k)$ in the full polarization tensor in the
sequel. The photon propagator at one-loop approximation is then given by
\beqa\label{eq:ph-propagator}
G_{\mu\nu}(k_0,\mathbf{k}) &=& \frac{-i P_{\mu\nu}}{k^2 - \pi_T(k_0,\mathbf{k})+ i\varepsilon} + 
\frac{-i Q_{\mu\nu}}{k^2 - \pi_L(k_0,\mathbf{k}) + i\varepsilon}
\eeqa
in the Feynman gauge, where $k^2 \equiv k_0^2 - \mathbf{k}^2$ by convention. 

\subsection{Spectral function and poles in photon propagator}
\label{subsec:sf-and-p}

We examine the influence of small black holes. The size of their
event horizon is extremely small, i.e. $r_H \ll 1.49{\times}10^{-14}\,\text{m}$. It means that
$g_R \ll 2.5{\times}10^{-14}$ for $R = 1\,\text{m}$ and, therefore, the one-loop correction to the photon
self-energy is small despite of $T_H$ is much larger than $0.5\,\text{MeV}$ or $6{\times}10^9\,\text{K}$.
Consequently, the photon dispersion relation approximately reads $k_0 \approx |\mathbf{k}|$.
The non-vanishing constant value of $\pi_T(k_0,\mathbf{k})$ in the limit $|\mathbf{k}| \rightarrow k_0$
implies, however, that the pole structure of the photon propagator is slightly modified, namely we now have
$k_0^2 = \mathbf{k}^2 + m_\gamma^2$ with $|\mathbf{k}| \gg m_\gamma$ (but still $T_H \gg |\mathbf{k}|$),
where the effective photon mass reads
\beqa
m_\gamma^2 &=& \lim_{|\mathbf{k}| \rightarrow k_0}\pi_T(k_0,\mathbf{k}) \;\approx\; \frac{1}{6}\,e^2T_L^2
\quad \text{with} \quad T_L \;\equiv\; \sqrt{g_R}\,T_H\,.
\eeqa
Thus, although we have employed the approximate expression for the fermion propagator
in~\cite{Emelyanov-16b}, we re-derive our main result of that paper by using the improved propagator
$S(x,x')$. It should also be mentioned that the local ($L$) temperature $T_L \rightarrow 0$ in the spatial
infinity unlike the Hawking temperature $T_H \neq 0$, because of $\sqrt{g_R} \propto r_H/R \rightarrow 0$
for $R \rightarrow \infty$.

The physical content of the poles appearing in the photon propagator \eqref{eq:ph-propagator} can 
be extracted by studying the analytic properties of the propagator~\cite{LeBellac}. We find that 
the spectral function $\rho(k_0,\mathbf{k})$ (equaling $2\pi\varepsilon(k_0)\delta(k^2)$ in the limit
$\alpha \rightarrow 0$, where $\alpha$ is the fine structure constant) is saturated by the transverse
pole, while the longitudinal pole gives a contribution that is of the order of $m_\gamma/|\mathbf{k}| \ll 1$.
This means that the transverse pole corresponds to the propagating mode, whereas the longitudinal
pole does not. It appears to be analogous to the behaviour of the transverse and longitudinal
mode (photon and plasmon, respectively) in the hot electron-positron plasma
for $e T \ll |\mathbf{k}| \ll T$~\cite{Weldon,LeBellac}.

\subsection{Screening of static electric field}
\label{subsec:ssef}

We now go over to the study of the electrostatic field $\mathbf{E} = -\nabla \varphi$ sourced by a
point-like charge $q$ in the presence of a small black hole. The electrostatic potential is given by
\beqa
\varphi(r) &=& q{\int}\frac{d^3\mathbf{k}}{(2\pi)^3}
\frac{\exp(i\mathbf{k}\mathbf{x})}{\mathbf{k}^2 + \pi_L(0,\mathbf{k})}
\quad \text{with} \quad 
\mathbf{k}\mathbf{x} \;=\; kr\cos\theta\,,
\eeqa
as this immediately follows from the linear response theory, where $\pi_L(0,\mathbf{k})$ must in
turn be computed in the limit $\beta|\mathbf{k}| \rightarrow 0$. We find
\beqa\label{eq:static-potential}
\varphi(r) &=& \frac{q}{4\pi^2}\int\limits_{0}^\infty dk k^2\int\limits_{0}^\pi d\cos\theta\,
\frac{\exp(i k r \cos\theta)}{k^2 - m_\gamma^2\tan^2\theta}\,.
\eeqa

To evaluate the integral in Eq.~\eqref{eq:static-potential}, we first expand the denominator of the integrand
over the parameter $m_\gamma^2/\cos^2\theta$ and then integrate it order by order over the angle
$\theta$.\footnote{The point $\theta = \pi/2$ is regular as follows from $\pi_L(0,\mathbf{k})$ for
$\theta = \pi/2$ and it does not contribute as can be directly shown.} Afterwards, we rewrite the integration
with respect to $|\mathbf{k}|$ to have it over $(-\infty,+\infty)$ (see Appendix \ref{app:electrostatic-potential}
for more details). This yields
\beqa\label{eq:esp}
\varphi(r) &\approx& \frac{q}{4\pi r}\,\exp(-r/r_D)
\quad \text{with} \quad r_D \;\equiv\; 1/(\gamma_L m_\gamma)\,.
\eeqa
Thus, we re-derive our result obtained in~\cite{Emelyanov-16b} by using the improved expression for
the fermion propagator, but with the Debye-like radius $r_D$ given by $(\gamma_L m_\gamma)^{-1}$
instead of $(\sqrt{2} m_\gamma)^{-1}$, where $\gamma_L$ appears to equal $\pi/2$ (see Appendix
\ref{app:electrostatic-potential}). This allows us to slightly enlarge the value of the maximal distance to
the small black hole which should still be ``visible" to a detector used in~\cite{Williams&Faller&Hill}
for testing the Coulomb law. Specifically, the small black hole should be in the region of the size about
$R_0 \approx 280\,\text{km}$ in order to discover the Debye-like screening of the electrostatic potential
induced by that.

It appears that we can even improve the estimate of $R_0$ to roughly one order of magnitude if we
take into account the correction $(m_\gamma r)^2\log(m_\gamma r)$ to the exponential function in
Eq.~\eqref{eq:esp} which is derived in Appendix \ref{app:electrostatic-potential}. Specifically, this 
correction leads approximately to the following modified Gauss law
\beqa
\Delta\varphi &\approx& -4\pi\rho + (\gamma_L m_\gamma)^2\varphi
+ 2m_\gamma^2\log(m_\gamma r)\varphi
\quad \text{for} \quad m_\gamma r \;\ll\; 1\,,
\eeqa
where $\rho$ is a charge density. Repeating computations of~\cite{Williams&Faller&Hill} with
this modified law, we obtain that $R_0 \approx 1.9{\times}10^{3}\,\text{km}$, where we have assumed
that the size of the conducting shells in~\cite{Williams&Faller&Hill} is about $1$ meter.

\subsection{Screening of static magnetic field}
\label{subsec:ssmf}

It turns out that there exists a local shielding effect for the magnetostatic field $\mathbf{B}$ as well. This
follows from the fact that $\pi_T(0,\mathbf{k}) \neq 0$ in the limit $|\mathbf{k}| \rightarrow 0$. This is
in sharp contrast to the normal hot plasma, wherein $\pi_T(0,\mathbf{k}) \rightarrow 0$ in that limit. It
should be mentioned that this effect does not exist for small \emph{eternal} black holes, because
$\pi_T(k_0,\mathbf{k})$ has the same structure as in the hot (isotropic) plasma and, hence, it vanishes
for $|\mathbf{k}| \rightarrow 0$.

As an example, we want to consider the screening of a static magnetic field $\mathbf{B}$ sourced by
the magnetic monopole of charge $q_m$. Introducing the magnetostatic potential $\varphi_m$, such
that $\mathbf{B} = - \nabla\varphi_m$, we find
\beqa
\varphi_m(r) &=& q_m{\int}\frac{d^3\mathbf{k}}{(2\pi)^3}
\frac{\exp(i\mathbf{k}\mathbf{x})}{\mathbf{k}^2 + \pi_T(0,\mathbf{k})}
\quad \text{with} \quad 
\mathbf{k}\mathbf{x} \;=\; kr\cos\theta\,.
\eeqa
Computing $\pi_T(0,\mathbf{k})$ in the limit $\beta |\mathbf{k}| \rightarrow 0$ and then
repeating the analysis of Sec.~\ref{subsec:ssef}, we obtain
\beqa
\varphi_m(r) &\approx& \frac{q_m}{4\pi r}\,\exp(-r/\bar{r}_D) \quad \text{with} \quad
\bar{r}_D \;\equiv\; \sqrt{2}/(\gamma_T m_\gamma)\,,
\eeqa
where $\gamma_T \approx 0.532818$ (see Appendix \ref{app:electrostatic-potential}). Thus, we find 
that $\bar{r}_D/r_D \approx 4$. It implies that the screening of the magnetostatic potential of the monopole
$q_m$ is more effective than that of the electrostatic potential of the charge $q$.

\section{Concluding remarks}

\subsection{Improved Wigner distribution}

We have derived the exact correction  to the Minkowski part of the propagator. This is non-singular and
induced by black holes in the far-horizon region. It is exact in that sense that this precisely satisfies
the field equation up to the terms vanishing faster than $1/R^2$ for $R \gg r_H$. Substituting this in the
definition of the Wigner distribution $\mathcal{W}(x,p)$~\cite{Emelyanov-17a}, we obtain
for the massless scalar field that
\beqa
\mathcal{W}(x,p) &=& \frac{1}{8\pi^2 p_0^3 R^2}\,\frac{\Gamma_{p_0}}{e^{\beta p_0} -1}\,
\delta(p_0 - p)\delta(p^\theta)\delta(p^\phi)\,,
\eeqa
where $\mathbf{p} = (p^r,p^\theta,p^\phi)$ and we have set $p^r \equiv p$. The parameter $\Gamma_{p_0}$
is given in Eq.~\eqref{eq:emt}. This implies that the effective Wigner distribution introduced
in~\cite{Emelyanov-17a} appears to be an exact result (up to the terms $1/R^n$ with $n \geq 3$).

\subsection{Quantum vacuum as anisotropic hot plasma}

We have found that there exists a local shielding effect for the magnetostatic field which is induced by
small black holes. The analogous effect can occur in the hot plasma which is described by the one-particle
distribution with the anisotropy in momentum space.

Although it is tempting to describe the local electromagnetic effects in the presence of small black
holes as if the vacuum is a plasma-like medium, this analogy seems to be incomplete. Indeed, this
``medium" cannot support the plasmon-like excitations which are normally attributed to the collective
excitations of the plasma particles~\cite{LeBellac}. Specifically, the plasma-like frequency
$\omega_p$ characterising these excitations can be computed by considering the limit
$|\mathbf{k}| \rightarrow 0$ with $k_0 \sim eT_L \ll eT_H$ in $\pi_T(k_0,\mathbf{k})$ and $\pi_L(k_0,\mathbf{k})$.
It turns out that $\omega_p$ for the transverse and longitudinal mode are different and depend on
the angle between $\mathbf{k}$ and the radial unit vector $\mathbf{n}$. We found in~\cite{Emelyanov-16b}
that the mode of the frequency $k_0 \sim eT_L$ has a wavelength which is much larger than the distance to
the black-hole centre $R$. This kind of waves cannot be described within our approximation. At these
scales, the hot-anisotropic-plasma analogy may not hold. 

\subsection{Modified dispersion relation of photon}

We found in~\cite{Emelyanov-16b} as well as in Sec.~\ref{subsec:sf-and-p} above that the photon dispersion
relation modifies in the presence of black holes, namely photons acquire a mass
term $m_\gamma$. In the far-horizon region, one has
\beqa
m_\gamma^2 &\propto& + \alpha T_L^2
\left\{
\begin{array}{cccc}
1\,, & T_H & \gg & m_e\,, \\[1mm]
(m_e/T_H)^\frac{3}{2}\exp(-m_e/T_H)\,, & T_H & \ll & m_e\,,
\end{array}
\right.
\eeqa
which vanishes when one neglects the interaction term between the electron/positron and
electromagnetic field.

In the near-horizon region, the effective photon mass squared
$m_\gamma^2$ might be \emph{negative}. Indeed, the polarization tensor can be computed
within the kinetic theory by employing the one-particle distribution function and the transport
equation. We found in~\cite{Emelyanov-17a} that the one-particle distribution near the event
horizon is negative. This might imply that photons can come out of the event
horizon~\cite{Emelyanov-17c}.\footnote{Note that the flux of these positive-energy photons
has a different nature in comparison with that of the Hawking radiation leading to the decrease
of the event-horizon size. The former is due to various quantum processes which might occur
in matter inside the horizon, whereas the latter is featureless and originates well outside black
holes. Thus, this kind of photons if existent could bring us information about  internal structure
of black holes.}

\section*{
ACKNOWLEDGMENTS}

It is a pleasure to thank Jos\'{e} Queiruga for discussions.

\begin{appendix}

\section{Scalar Feynman propagator}
\label{app:feynman-propagator}

We have derived a correction to the Minkowski 2-point function in the far-horizon region for a massless
scalar field in~\cite{Emelyanov-17a}. This correction can be written as follows
\beqa
\Delta{W}(x,x') &\approx& +g_R{\int}\frac{d^3\mathbf{k}}{(2\pi)^3}\,\frac{n_\beta(k_0)}{k_0}\,
\exp(i\mathbf{k}\Delta\mathbf{x})\Big(1-\frac{i}{2}\mathbf{k}\Delta\mathbf{x}\Big)
\cos\big(\bar{k}\Delta{x}\big)\,.
\eeqa
where $\bar{k}^\mu \equiv k_0(1,\mathbf{n})$ by definition and we have omitted cubic- and higher-order
terms with respect to $\Delta\mathbf{x}$ as well as those terms which vanish faster than $1/R^2$ in
the asymptotically flat region.

The correction to the scalar Feynman propagator is thus given by
\beqa
\Delta{G}(x,x') &\approx& +2\pi g_R{\int}\frac{d^4k}{(2\pi)^4}\,\frac{\delta(k^2)}{e^{\beta|k_0|} - 1}\,
\Big(1-\frac{i}{2}\mathbf{k}\Delta\mathbf{x}\Big)\exp(-i\bar{k}\Delta{x} + i\mathbf{k}\Delta\mathbf{x})\,.
\eeqa
Bearing in mind the structure of the radial modes, we want to find a function
$h \equiv h(i\mathbf{k}\Delta\mathbf{x})$ which satisfies the following conditions
\bsubeqs
\beqa
h &=& 1- \frac{i}{2}\mathbf{k}\Delta\mathbf{x} + \text{O}\big((\mathbf{k}\Delta\mathbf{x})^2\big)\,,
\\[1mm]
0 &=& h^{\prime\prime} + 2h^\prime + h\,,
\eeqa
\esubeqs
where the prime denotes the differentiation with respect to the argument of the function $h$.
The second condition implies that $\Delta{G}(x,x')$ is a solution of the scalar field equation, i.e.
$\Box \Delta{G}(x,x') = 0$, up to the terms vanishing faster than $1/R^2$ for the large values of $R$.
Thus, we obtain
\beqa
\Delta{G}(x,x') &\approx& +2\pi g_R{\int}\frac{d^4k}{(2\pi)^4}\,\frac{\delta(k^2)}{e^{\beta|k_0|} - 1}\,
\exp(-i\bar{k}\Delta{x})\,.
\eeqa
This result can be directly employed to derive $\Delta{S}(x,x')$ for the massless Dirac field.

\section{Computation of electrostatic potential}
\label{app:electrostatic-potential}

The electrostatic potential we compute here reads
\beqa
\varphi(r) &=& \frac{q}{4\pi^2}\sum\limits_{n = 0}^{+\infty}\int\limits_{0}^\infty dk\,
\frac{k^2m_\gamma^{2n}}{(k^2 + m_\gamma^2)^{n+1}}\int\limits_{-1}^{+1} dz\,
\frac{\exp(i k r z)}{z^{2n}} \;\equiv\; \sum\limits_{n = 0}^{+\infty} \varphi_n(r)\,.
\eeqa
We first consider the term $n = 0$. One has
\beqa
\varphi_0(r) &=& \frac{i q}{4\pi^2} \int\limits_{-\infty}^{+\infty} dk\,\frac{k\, e^{-ikr}}{k^2 + m_\gamma^2} 
\;=\; \frac{q}{4\pi r}\,\exp(-m_\gamma r)\,,
\eeqa
where we have chosen the contour $C_\infty$ to evaluate the integral over $k$ by employing the residue
theorem. This contour is depicted in Fig.~1.
\begin{figure}[t]
\includegraphics[width=11.0cm]{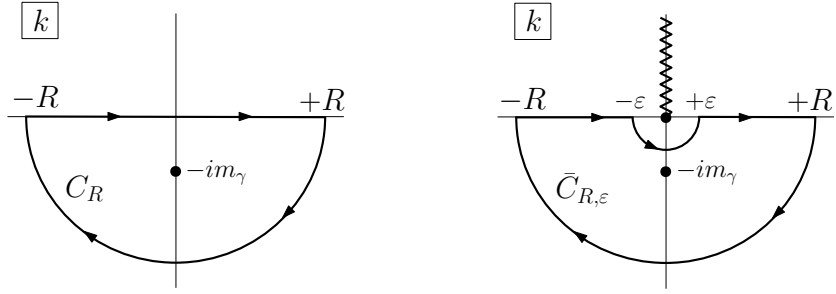}
\caption{The contours $C_\infty$ and $\bar{C}_{\infty,0}$ are chosen when the integrand contains the
exponential function and the exponential integral function, respectively. The residue theorem is then
used to evaluate the principal value integrals over the real values of $k$.}
\end{figure}
The next term in the expansion of the potential $\varphi(r)$ reads
\beqa
\varphi_1(r) &=& \frac{q m_\gamma^2}{4\pi^2}\int\limits_{-\infty}^{+\infty} dk\,
\frac{irk^3 \text{E}_1(ikr) -k^2e^{-ikr}}{(k^2 + m_\gamma^2)^2} \;\approx\; \frac{q}{4\pi r}(-m_\gamma r/2)
\quad \text{for} \quad m_\gamma r \;\ll\; 1\,,
\eeqa
where we have evaluated the integral with the exponential integral with the complex
argument~\cite{Abramowitz&Stegun}, $\text{E}_1(z)$, by choosing the contour $\bar{C}_{\infty,0}$
shown in Fig.~1. Employing this procedure for higher values of $n$, we obtain
\beqa
\varphi(r) &\approx& \frac{q}{4\pi r}\,\exp(-\gamma_L m_\gamma r)
\Big(1 + (m_\gamma r)^2\ln(m_\gamma r)\Big)
\quad \text{for} \quad m_\gamma r \;\ll\; 1\,,
\eeqa
where by definition
\beqa\label{eq:gl}
\gamma_L &\equiv& 1 + \frac{1}{2} + \frac{1}{3{\cdot}2^3} + \frac{1}{5{\cdot}2^4}
+ \frac{5}{7{\cdot}2^7} + \frac{7}{9{\cdot}2^8} + \frac{3{\cdot}7}{11{\cdot}2^{10}}
+ \frac{3{\cdot}11}{13{\cdot}2^{11}} + \cdots \;\approx\; 1.570051\,,
\eeqa
where we have taken into account the first $26$ terms in the series. Since $\pi/2 \approx 1.570796$, we 
conjecture that $\gamma_L = \pi/2$ exactly. For a later use, we also define
\beqa\label{eq:gt}
\gamma_T &\equiv& 1 - \frac{1}{2} + \frac{1}{3{\cdot}2^3} - \frac{1}{5{\cdot}2^4}
+ \frac{5}{7{\cdot}2^7} - \frac{7}{9{\cdot}2^8} + \frac{3{\cdot}7}{11{\cdot}2^{10}}
- \frac{3{\cdot}11}{13{\cdot}2^{11}} + \cdots \;\approx\; 0.532818\,
\eeqa
that holds for the first $26$ terms in the series.

\end{appendix}

\end{document}